# Combinatorial approach to bulk detector material engineering: Application to rapid NaI performance optimization via multi-element doping/co-doping strategy

I. V. Khodyuk[a], S. A. Messina, T. J. Hayden, E. D. Bourret, G. A. Bizarri

*Lawrence Berkeley National Laboratory, 1 Cyclotron Road, Berkeley, CA 94720, USA.*

*Historically, the discovery and optimization of doped bulk materials has been predominantly developed through an Edisonian approach. While successful and despite the constant progress in fundamental understanding of detector materials physics, the process has been restricted by its inherent slow pace and low success rate. This poor throughput owes largely to the considerable compositional space that needs to be accounted for to fully comprehend complex material/performance relationship. Here, we present a combinatorial approach where doped bulk scintillator materials can be rapidly optimized for their properties through concurrent extrinsic doping/co-doping strategies. The concept that makes use of Design of Experiment, rapid growth and evaluation techniques, and multivariable regression analysis, has been successfully applied to the engineering of NaI performance, a historical but mediocre performer in scintillation detection. Using this approach, we identified a three-element doping/co-doping strategy that significantly improves the material performance. The composition was uncovered by simultaneously screening for a beneficial co-dopant ion among the alkaline earth metal family and by optimizing its concentration and that of $Tl^+$ and $Eu^{2+}$ ions. The composition with the best performance was identified as 0.1% mol $Tl^+$, 0.1% mol $Eu^{2+}$ and 0.2% mol $Ca^{2+}$. This formulation shows enhancement of energy resolution and light output at 662 keV, from 6.3 to 4.9%, and from 44,000 to 52,000 ph/MeV, respectively. The method, in addition to improving NaI performance, provides a versatile framework for rapidly unveiling complex and concealed correlations between material composition and performance, and should be broadly applicable to optimization of other material properties.*

The discovery and optimization of multi-element compounds out of a large combinatorial space is a daunting task. It has been especially challenging for doped bulk materials where one has to account for concentrations ranging over several orders of magnitude, from elemental composition (lattice) to ppm levels (dopants). This large compositional space has definitely challenged and slow downed the development of the next generation of scintillator materials, where despite an increasing theoretical understanding of the material/performance relationship, the process is predominantly developed through a time-consuming Edisonian approach[1]. Even for relatively simple binary systems, computational techniques are still falling short of fully comprehending the complex interplay between composition, energy flow and material performance. While the use of combinatorial approach in order to account for large parameter space has been profitable for thin film[2] and powder forms material development[3], it has only been marginally successful when applied to bulk material. The difficulties to rapidly synthesize single crystal materials and to measure representative bulk properties have always impeded the extraction of clear trend or patterns.

We present here a combinatorial approach in which doped bulk scintillator materials can be optimized for their properties through concurrent extrinsic doping/co-doping strategies. To our knowledge this is the first attempt to employ a combinatorial approach for detector materials discovery and optimization. The concept relies on a three-step process: (i) experimental planning and the application of Design of Experiment (DoE), (ii) material synthesis and characterization with the use of rapid single crystal growth and evaluation techniques, and (iii) data analysis leveraging response surface and multivariable regression analysis methods. The core of the Design of Experiment revolves around a Taguchi method[4]. The Taguchi framework is particularly well adapted to simultaneously study multiple factors influence on a targeted output parameter. The arrangement of the experimental set, an orthogonal array[5,6], is designed to explore and optimize the material performance in a multi-dimensional space using the least possible number of experiments. This framework was coupled to the LBNL high-throughput synthesis and characterization facility[7] to rapidly produce and evaluate single crystalline samples based on a non-directional solidification technique[8].

[a] Correspondence should be addressed to I.V. Khodyuk. Electronic mail: ivkhodyuk@lbl.gov.

The entire approach was applied to NaI:Tl$^+$ and engineering of its performance, both energy resolution and light output wise. The choice of NaI:Tl$^+$ was twofold: (i) its importance for the scintillation field, and (ii) the long-lasting scientific challenge to understand, control and improve its performance. NaI:Tl$^+$ performance is considered as mediocre with a moderate light output (LO) of 44,000 photons/MeV and a poor energy resolution (ER) of at best 6.3% at 662 keV[9]. Improving its performance has been an important scientific challenge for the scintillation community. Most of the efforts have been directed toward crystal growth process optimization and/or extrinsic element addition to the melt. Most of the elements were found to be at most transparent to NaI:Tl$^+$ performance. Those included Mn, Pb, Ag, oxides, chalcogens, and halogens at low concentrations[10]. To our knowledge, the best published results of NaI performance are from Shiran *et al.* where adding Eu$^{2+}$ to NaI:Tl$^+$ showed an improvement of the light output (48,000 photons/MeV) and of the energy resolution (6.2%)[11]. Recently, there has been a renewed interest among the scintillator community to revisit co-doping strategies. The main driving force was the successful demonstration that co-doping LaBr$_3$:Ce$^{3+}$ with 200 ppm of Sr$^{2+}$ considerably improves the material energy resolution, from 2.7% to 2.0% at 662 keV[12]

*Table 1 – Orthogonal experimental set composition and characterization results.*

| Design # | [Tl$^+$] (mol %) | IIA | [IIA$^{2+}$] (mol %) | [Eu$^{2+}$] (mol %) | Light output @ 662keV (10$^3$ ph/MeV) | Energy resolution @ 662keV (%) | $\lambda_{emission}$ (nm) | $\Delta\lambda_{FWHM}$ |
|---|---|---|---|---|---|---|---|---|
| 0 | 0.1 | - | 0 | 0 | 43 | 7.0 | 419 | 116 |
| 1 | 0.0 | Mg | 0.1 | 1.0 | 24.6 | 8.5 | 473 | 34 |
| 2 | 0.0 | Ca | 0.2 | 0.5 | 37.3 | 6.4 | 473 | 34 |
| 3 | 0.0 | Sr | 0.4 | 0.1 | 14.5 | 9.9 | 463 | 37 |
| 4 | 0.0 | Ba | 0.8 | 0.0 | 5.1 | 21 | 320 | 140 |
| 5 | 0.1 | Mg | 0.2 | 0.0 | 18.3 | 13.4 | 417 | 112 |
| 6 | 0.1 | Ca | 0.1 | 0.1 | 41.6 | 6.9 | 465 | 43 |
| 7 | 0.1 | Sr | 0.8 | 0.5 | 35.5 | 8 | 470 | 34 |
| 8 | 0.1 | Ba | 0.4 | 1.0 | 4.1 | 13.1 | 474 | 34 |
| 9 | 0.25 | Mg | 0.4 | 0.5 | 12.4 | 20 | 473 | 34 |
| 10 | 0.25 | Ca | 0.8 | 1.0 | 33.9 | 7 | 468 | 36 |
| 11 | 0.25 | Sr | 0.1 | 0.0 | 29.9 | 6.1 | 418 | 115 |
| 12 | 0.25 | Ba | 0.2 | 0.1 | 47 | 5.9 | 447 | 26 |
| 13 | 0.5 | Mg | 0.8 | 0.1 | 33.4 | 17.5 | 447 | 29 |
| 14 | 0.5 | Ca | 0.4 | 0.0 | 22.6 | 10.9 | 452 | 101 |
| 15 | 0.5 | Sr | 0.2 | 1.0 | 23.8 | 12 | 475 | 34 |
| 16 | 0.5 | Ba | 0.1 | 0.5 | 16.7 | 17.5 | 472 | 41 |

These endeavors enticed us to revisit the engineering of NaI using multi-element doping/co-doping strategy. The experimental planning, largely driven by the studies summarized here before, was devised to simultaneously study and optimize NaI energy resolution and light output at 662 keV as a function of the Tl$^+$ concentration ([Tl$^+$]), the addition of a co-dopant ion chosen among alkaline earth metal family (type, IIA and concentration [IIA$^{2+}$]), and the concentration of a second emitting center, europium ([Eu$^{2+}$]). Leveraging the work of Taguchi[4], the compositional space was discovered through experimental set organized to form a L$_{16}$ orthogonal array (4 levels per parameter, also called factors – Tables 1). This arrangement that can be classified as fractional factorial design[13] allows for surveying the main effect of the factors on the targeted objectives while drastically reducing the number of required experiments. A 4-factors/4 levels full factorial design will require an unpractical set of 256-experiments to cover the same combinatorial space. A reference sample, NaI:Tl$^+$ doped with 0.1 mol % Tl$^+$, was included in the experimental set for control and comparison purposes. All the concentrations are given in mole percent and correspond to the nominal concentration of the starting materials.

The 17 samples were synthesized following a non-directional solidification approach using 5N pure anhydrous beads of NaI, MgI$_2$, CaI$_2$, SrI$_2$, BaI$_2$, TlI and EuI$_2$ from Sigma-Aldrich as starting material. The sample preparation, weigh and ampoule encapsulation was done in an argon-filled drybox maintained below 0.1 ppm of O$_2$ and H$_2$O. The ampoules were then sealed under dynamic vacuum pumping and placed in a horizontal furnace, heated to 675ºC to melt the NaI and held at this temperature for 6 hours, in order to homogenize the liquefied contents of the quartz ampoule. After solidification all samples were transferred back inside the dry box and processed for characterization in form of powder for x-ray diffraction (XRD), 10 mm$^3$ single crystalline pieces for pulse height measurements (PHM) and airtight quartz cuvettes filled with 0.5-2 mm$^3$ crystal pieces for x-ray luminescence (XRL) measurements.

The correct crystal structure phase of each sample was confirmed measuring their powder XRD patterns[7]. A minimum of 3 crystal pieces per composition were selected for pulse-height measurements under $^{137}$Cs excitation. The photopeak centroid and full-width at half-maximum were determined by using a superposition of two Gaussian functions for the photopeak and an x-ray escape peak, and of an exponential background[14]. The light output was estimated by comparison of the photopeak position of the sample of interest with the response of a 10 mm$^3$ commercial NaI:Tl$^+$ crystal from ScintiTech[15] measured under identical conditions.

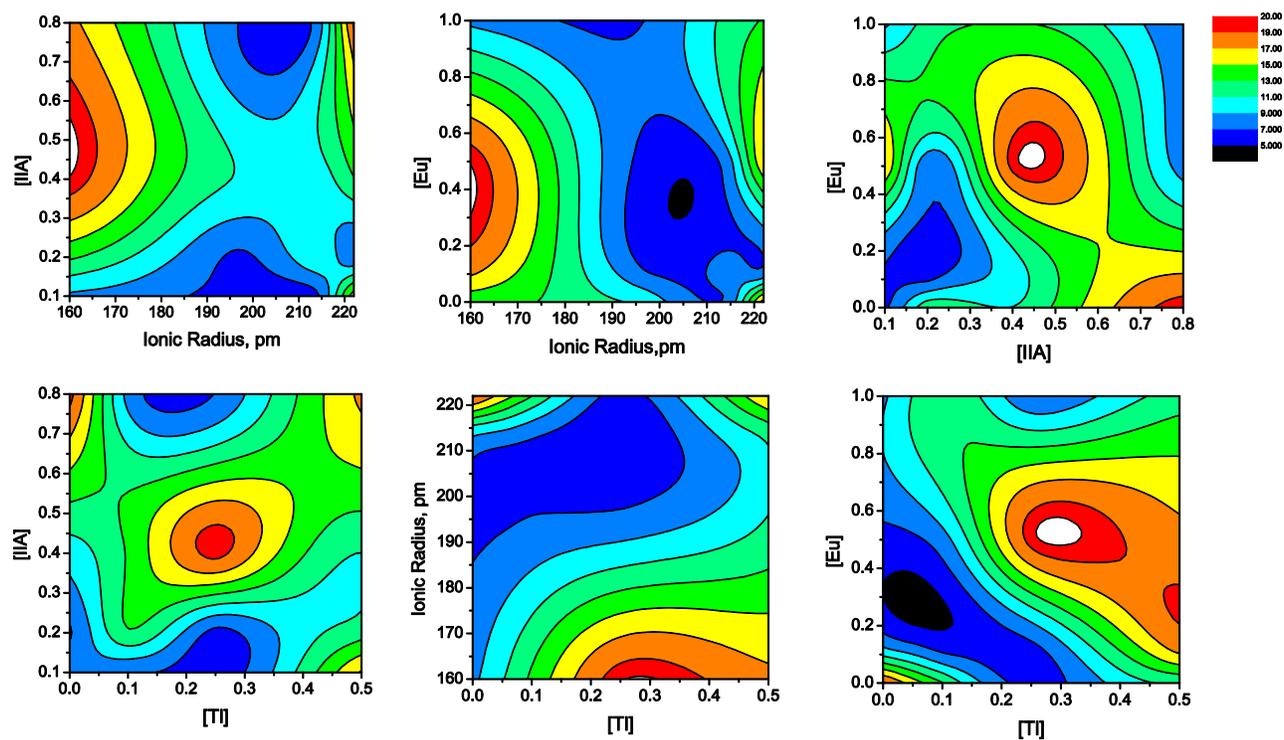

*Fig. 1 - Response surfaces of NaI:Tl, X, Y samples energy resolution as a function of [Tl$^+$], [Eu$^{2+}$], [IIA$^{2+}$] and IIA type. To plot response surfaces IIA element types were substituted with corresponding ionic radii in pm: Mg – 160, Ca – 197, Sr – 215 and Ba – 222.*

The 2D response maps were determined based on the results from Table 1 using the DOE.base package from the language R[16] and Qualiteck-4[5] software. The maps allow for an estimation of which explanatory factors have an impact on the light output and energy resolution as well as a determination of which compositional set gives the optimal response within the combinatorial space explored. To bypass the non-mathematic formulation and inherent granularity of the factor co-dopant ion type, we substitute the factor ion type by its associated ionic radius in pm. The maps for the energy resolution are presented in Fig. 1. For the LO and the ER, the optimal response coincided with the composition 0.25 mol % $Tl^+$, 0.2 mol % $Ca^{2+}$ and 0.1 mol % $Eu^{2+}$.

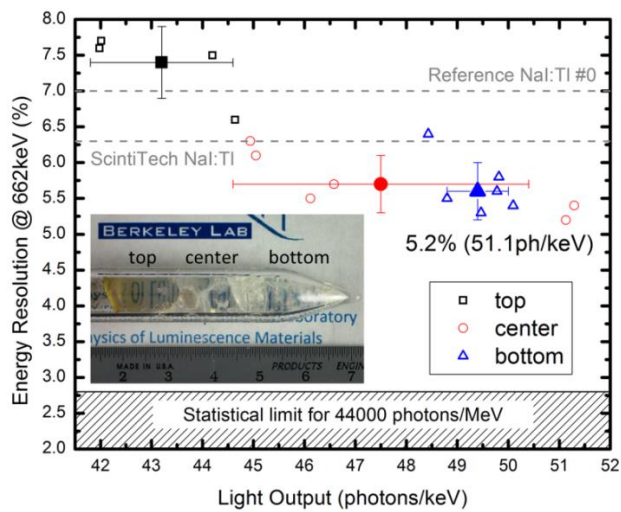

*Fig. 2 - Energy resolution versus light output of NaI:0.25%$Tl^+$,0.2%$Ca^{2+}$, 0.1%$Eu^{2+}$. Inset: Picture of crystal rapidly grown in quartz ampoule.*

To test the validity of the multivariable regression analysis output, two additional samples were synthesized with the optimal composition formula. The first sample was synthesized using the analogous non-directional solidification approach while the second one was grown using a conventional vertical Bridgman-Stockbarger technique. For the latter, the reactants were heated at about 200 ºC under vacuum to remove residual moisture. The growth was conducted in a sealed ampoule suspended in a vertical Bridgman furnace and translated through a thermal gradient of 10 ºC/cm at a rate of 0.8-2.0 mm/hour. The crystal was 10 mm in diameter and about 6 cm long.

Both samples were characterized for their light output and energy resolution at 662keV. The first sample gave the best results among the non-directional solidification sample set with a LO of 48,200 ph/MeV, and ER of 5.4%. For the Bridgman grown sample, measurements were taken on several samples collected along the direction growth axis, bottom, middle and top.

*Table 2 - Concentration of $Tl^+$, $Ca^{2+}$ and $Eu^{2+}$ in NaI lattice according to ICP-MS. (ADD mol %)*

| Position in the boule | [$Tl^+$] | | [$Ca^{2+}$] | | [$Eu^{2+}$] | |
|---|---|---|---|---|---|---|
| | (ppm wt) | (mol %) | (ppm wt) | (mol %) | (ppm wt) | (mol %) |
| Nominal in melt | 3470 | 0.25 | 540 | 0.20 | 1000 | 0.10 |
| Top | 14641 | 1.05 | 580 | 0.21 | 890 | 0.09 |
| Center | 1500 | 0.11 | 490 | 0.18 | 940 | 0.09 |
| Bottom | 880 | 0.06 | 580 | 0.21 | 940 | 0.09 |

The light output and energy resolution values as a function of the position along the growth direction are presented in Fig. 2. There emerged a significant difference in LO and ER between the top, center, and bottom parts of the crystal. The absolute best results were obtained for two crystals selected from the center part with ER of 5.2% and LO of 51,100 ph/MeV. In term of average values, the crystals selected from the bottom part shows a better uniformity in their response. Variation of the scintillation performance was expected due to the different segregation coefficients of the dopants and co-dopant. Dopants and co-dopant segregations can lead to a significant non-uniformity of their concentration distribution along the crystal. While the approach succeed to underline the beneficial pattern of using $Tl^+$, $Eu^{2+}$ and $Ca^{2+}$ as a set, the approach lacks of accuracy when it comes to quantify the optimum concentration of highly segregating elements. The level, nominal concentration of the elements is not descriptive enough and leads to loosen the constraint imposed by the data set on the output of the multivariable regression analysis.

To determine actual concentration of the elements along the growth axis inductively coupled plasma mass spectrometry has been done. As shown on the quantitative elemental distribution of the $Tl^+$, $Ca^{2+}$ and $Eu^{2+}$ (Table 2), thallium is heavily segregated during the growth. This is clearly noticeable on the picture of the crystal presented in the inset of Fig. 2 where a clear yellow layer, corresponding to a high thallium concentration area, is visible at the top of the boule. However, there is only minor inhomogeneity in $Ca^{2+}$ and $Eu^{2+}$ distribution throughout the crystal.

To better quantify the optimal thallium concentration, a second crystal with 0.1 mol % $Tl^+$ was grown using the same Bridgman-Stockbarger technique. Single crystalline pieces from different parts of the crystal show LO above 50,000 ph/MeV and ER around 5.0% at 662keV. The best light output of 52,000 ph/MeV and an energy resolution of 4.9% at 662 keV were recorded for one of the crystals from the middle part of the boule. The photopeak from this crystal is shown in Fig.3 in comparison with the commercial and homemade reference samples.

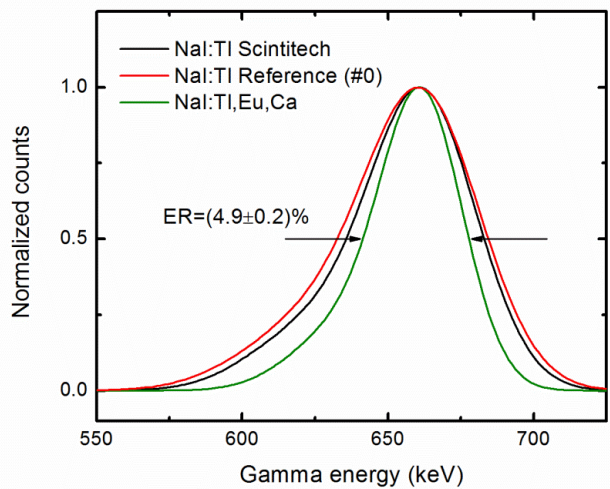

*Fig. 3 - Pulse-height spectra of $^{137}Cs$ recorded with commercial NaI:Tl and NaI doped with 0.1%$Tl^+$, 0.2%$Ca^{2+}$ and 0.1%$Eu^{2+}$.*

In this study, we present a combinatorial approach allowing to rapidly exploring the relationships between material composition and material properties. To the authors' knowledge it is the first report of the application of this technique to the optimization of doped bulk scintillators performance. The approach is particularly adapted to guide the development of detector and luminescent materials where the compositional landscape becomes more and more complex due to a large number of variables and/or complex interdependencies of the factors as well as the extremely demanding level of performance required.

The approach was successfully applied to the optimization of the light output and energy resolution of NaI as a function of multiple elements doping/co-doping strategies. The results show a drastic improvement of both properties. Optimized sample shows an improvement of its energy resolution down to 4.9% at 662 keV and a light output up to 52,000 ph/MeV. It is expected that the performance of NaI can still be improved by narrowing compositional space toward the optimal composition and by improving the crystal growth process and purification of the starting materials. The literature gives indication of the potential intrinsic value that can be reached with NaI. Undoped NaI light output at 662 keV has been reported[17] above 80,000 ph/MeV when measured at liquid nitrogen temperature with an energy resolution of about 4%.

Finally, it is reasonable to think that this combinatorial approach can be extended to other objectives and/or factors. For the latter, variables such as material stoichiometry and/or growth parameters (temperature gradient, atmosphere etc) are certainly a logical extension of the combinatorial space. Comparably targeting the optimization of other detector requirements such as minimization of the self-absorption or maximization of the neutron/gamma-ray discrimination in dual modality detectors is also a coherent direction of this work. However it is important to stress out two pivotal axioms of the approach: (i) Data cannot be collected without some preexisting ideas about what may or may not be relevant to the specific problem. There is no mathematical expression telling which particular factors must be examined in a given study. In our case, the decision was heavily driven by former experimental and theoretical studies. (ii) While the statistical analysis can identify patterns in complex experimental data sets, it does not provide what one makes of the pattern once discerned. In other word, once the data relevant to the problem are collected and the pattern underlined, how should the finding be explained and incorporated to a microscopic model of the phenomena is not a purpose of the approach.

**Acknowledgements**

The authors would like to thank S. Hanrahan, D. Wilson, and J. Powell for their technical and engineering support. Fruitful scientific discussions with Dr. G. Gundiah, Dr. M. Gascon, Dr. E. Samulon, Dr. D. Perrodin and Dr. S. Derenzo are highly appreciated. This work was supported by the US Department of Homeland Security/DNDO and the US Department of Energy/NNSA/NA22 and carried out at Lawrence Berkeley National Laboratory under Contract no. AC02-05CH11231. This work does not constitute an express or implied endorsement on the part of the government.